\documentclass[aps, reprint, prb, a4paper]{revtex4-1}
\usepackage{amsmath}
\usepackage{graphicx}
\usepackage{epstopdf}
\usepackage{subcaption}
\usepackage{bm}
\usepackage{epstopdf}
\usepackage{xcolor}

\newcommand{\beq}{\begin{equation}}
\newcommand{\eeq}{\end{equation}}
\newcommand{\ali}[1]{\begin{align} #1 \end{align}}
\newcommand{\spl}[1]{\begin{split} #1 \end{split}}
\newcommand{\ve}{\varepsilon}
\newcommand{\fe}{\mathcal{F}}
\DeclareMathOperator{\tr}{Tr}

\newcommand{\bpm}{\begin{pmatrix}}
\newcommand{\epm}{\end{pmatrix}}

\begin{document}

\title{Signatures of Nematic Superconductivity in Doped Bi$_2$Se$_3$ under Applied Stress}

\date{\today}
\author{Pye Ton How}
\affiliation{Institute of Physics, Academia Sinica, Taipei 115,
	Taiwan}

\author{Sung-Kit Yip}
\affiliation{Institute of Physics, Academia Sinica, Taipei 115,
	Taiwan}
\affiliation{Institute of Atomic and Molecular Sciences, Academia
	Sinica, Taipei 106, Taiwan}
\affiliation{Physics Division, National Center for Theoretical Sciences, Hsinchu 300, Taiwan}

\begin{abstract}
The $M_x \text{Bi}_2 \text{Se}_3$ family are candidates for topological superconductors, where $M$ could be Cu, Sr, or Nb.  Two-fold anisotropy has been observed in various experiments, prompting the interpretation that the superconducting state is nematic.  However, it has since been recognized in the literature that a two-fold anisotropy in the upper critical field $H_{c2}$ is incompatible with the na\"{i}ve nematic hypothesis.  In this paper we study the Ginzburg-Landau theory of a nematic order parameter coupled with an applied stress, and classify possible phase diagrams.  Assuming that the $H_{c2}$ puzzle is explained by a pre-existing "pinning field", we indicate how a stress can be applied to probe an extended region of the phase diagram, and verify if the superconducting order parameter is indeed nematic.  We also explore the Josephson tunneling between the proposed nematic superconducting state and an s-wave superconductor.  The externally applied stress is predicted to serve as an on/off switch to the tunneling current, and in certain regime the temperature dependence of the critical current can be markedly different from that between two conventional s-wave superconductors.
\end{abstract}

\maketitle

\section{Introduction}

$\text{Bi}_2 \text{Se}_3$ is a topological insulator \cite{Xia2009, Zhang2009, Liu2010}.  When intercalated with electron-donating atoms such as Cu\cite{Hor2010, Kriener2011}, Sr\cite{Shruti2015} or Nb\cite{Qiu2015, Asaba2017}, it becomes superconducting at low temperature.  Following their identification as candidates of topological superconductors \cite{Fu2010, Fu2014, Venderbos2016}, the family of materials has recently garnered a lot of interest\cite{Yonezawa2018}.  Experimentally, two-fold anisotropy in the superconducting phase has been observed in NMR Knight's shift\cite{Matano2016}, specific heat\cite{Yonezawa2017,Willa2018,Sun2019}, magnetic torque\cite{Asaba2017}, magneto-transport and upper critical field\cite{Pan2016,Yonezawa2017,Du2017,Smylie2018}.  This is incompatible with the crystal lattice symmetry, and has led to the hypothesis that the superconducting state is nematic in nature, i.e. it spontaneously breaks both lattice rotational and gauge symmetries.  While the superconductivity is suppressed by an applied hydrostatic pressure\cite{Bay2012,Nikitin2016}, the two-fold anisotropy is observed as far as superconductivity holds\cite{Nikitin2016}.

\begin{figure}
	\includegraphics[width=0.45\textwidth]{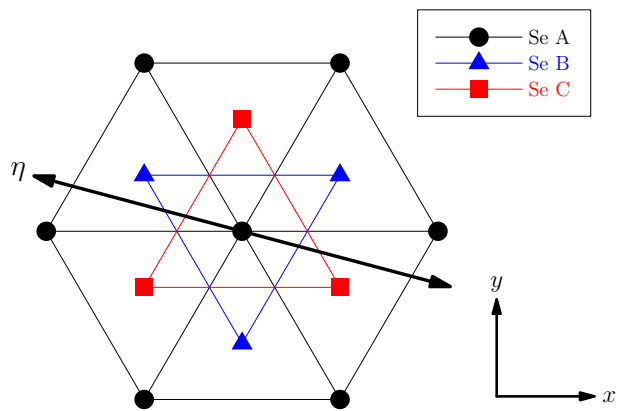}
	\caption{\label{quintuple} This shows the positions of the selenium atoms within a quintuple layer of $\text{Bi}_2\text{Se}_3$.  The superconducting order parameter $\vec{\eta}$ is shown here pointing in an arbitrary direction.}
\end{figure}

The parent material $\text{Bi}_2 \text{Se}_3$ forms a rhombohedral crystal, with a lattice point group $D_{3d}$; see FIG \ref{quintuple}.  According to the nematic hypothesis, the complex superconducting order parameter $\vec{\eta}$ is believed to transform under the irreducible representation $E_u$\cite{Fu2014}, which is parity-odd and two-dimensional, and, in the absence of other symmetry breaking effect, spontaneously breaks the crystal rotation symmetry together with the $U(1)$ symmetry.  It can be thought of as a nematic director in the basal plane, since $\vec{\eta}$ and $-\vec{\eta}$ can be identified up to a global $U(1)$ phase shift of $\pi$.  Experiments that detect the anisotropy in the basal plane all reported a two-fold symmetry, consistent with the presence of a nematic superconducting order parameter.  This na\"{i}ve picture, however, is contradicted by several pieces of experimental facts.

First of all, the upper critical field $H_{c2}$ is experimentally observed to be \emph{two-fold} anisotropic\cite{Pan2016,Yonezawa2017,Du2017,Smylie2018,Sun2019} with the applied field parallel to the basal plane, seemingly inline with other measurements.  However, this result is actually incompatible with the na\"{i}ve nematic hypothesis, which indicates a \emph{six-fold} anisotropy for $H_{c2}$\cite{Agterberg1995a, Krotkov2002, Venderbos2016a}.

To understand this, one considers the \emph{onset} of superconductivity in high field.  Instead of a finite $\vec{\eta}$ providing a preferred orientation, the magnetic field and the lattice anisotropy together determine the orientation of the infinitesimal $\vec{\eta}$ when it first emerges.  This consideration leads to the prediction of a six-fold anisotropy associated with the six-fold improper rotation symmetry of $D_{3d}$.

%{\color{red}

The robustness of the two-fold nematic direction observed in experiments is another puzzle unexplained by the na\"{i}ve nematic scenario.  It was noted that the two-fold anisotropy for a given sample is always pinned at the same direction.  In Pan \emph{et. al.}\cite{Pan2016} part of the experiment was performed with the sample repeatedly cycled up to $5$K (about $1.6 T_c$) and cooled back down to superconducting phase again before every field sweep\footnote{A. de Visser, private communication}.  In a separate account\footnote{M. P. Smylie, private communication}, the same two-fold direction for any given sample persists as it was cycled in and out of the superconducting phase repeatedly, sometimes all the way to room temperature and back, with or without a background magnetic field.  To account for this behavior, the rotational symmetry must have been already broken in some way at the room temperature.

Indeed, for the case of Sr, evidence of normal-state nematic response was reported by Kuntsevich \emph{et. al.}\cite{Kuntsevich2018} and Sun \emph{et. al.}\cite{Sun2019}.  On the other hand, the Argonne group\cite{Willa2018, Smylie2018} reported isotropic normal-state response for Sr.  This is in contrast with the case of Cu, where no nematicity was ever observed above $T_c$\cite{Matano2016,Yonezawa2017} to our knowledge.  

To reconcile the $E_u$ pairing scenario and the observed robust pinning effect, some nematic symmetry-breaking field (SBF) must already be present in the sample\cite{Willa2018}.  The precise nature of this SBF is yet unknown, though mechanical stress or strain were named as possible candidates \cite{Fu2014}.  This SBF pins all the two-fold nematic responses (including $H_{c2}$) in the same direction for a given sample.

We emphasize that any scenario that solely relies on spontaneous symmetry breaking at low temperature to supply the nematic direction is incompatible with experiments.  The na\"{i}ve nematic scenario does not work, nor does any scenario where the rotational symmetry is broken independent of the $U(1)$ symmetry, for example\cite{Hecker2018}.  A pre-existing SBF (``pining field'') is necessary.

A conceptual difficulty then arises: in the presence of a pinning field, an otherwise isotropic s-wave superconductor can in principle also appear two-fold anisotropic.  Can we tell this so-called anisotropic s-wave scenario apart from the true nematic scenario?  This distinction is more than merely academic, because the nematic superconductor is topological\cite{Fu2014, Venderbos2016}, while the s-wave case is not.

In light of these recent developments, a better understanding of how the proposed nematic superconductivity interacts with an explicit SBF is clearly desirable.  In this work we consider mechanical stress or strain as the prototypical SBF, although an SBF with a different microscopic nature will have exactly the same phenomenology.  We further consider the application of stress in addition to the existing, unknown SBF, as an experimental probe of the nature of the superconducting phase.
%}

Historically, much of the phenomenology of a superconducting order parameter with non-trivial symmetry has been known as part of the lore of the unconventional superconductors (see \cite{Sigrist1991, Joynt2002} and references therein.)  Pertinent to the present discussion is the fact that, due to the competition of the SBF and the crystal field, the superconducting transition may be split into two if the order parameter is multi-component\cite{Hess1989, Machida1989}.  In the heavy fermion compound $\text{UPt}_3$, the double transition is due to the antiferromagnetic order in the normal state, and has been experimentally observed\cite{Fisher1989}.  The idea of a stress causing similar split has been put forward for $\text{Sr}_2\text{Ru}\text{O}_4$\cite{Sigrist1987a, Walker2002}, but to our knowledge has never been observed\cite{Hicks2014,Watson2018}.

In this paper, we formulate the Ginzburg-Landau (GL) theory of a nematic superconducting order parameter coupled to an external SBF, and explore the phenomenology as a guide to future experiments.  In particularly, following the suggestion of \cite{Fu2014}, we consider stress as the prototypical SBF.  The GL theory is written down in Section \ref{GLsection}, and then in Sections \ref{PhaseDiagramSection}, \ref{SplitTransitionSection} and \ref{LowerTransitionSection}, we identify the splitting of the superconducting transition, as well as another possible transition at a lower temperature, and classify possible phase diagrams as function of stress and temperature.  In section \ref{PinningFieldSection}, we further explore the phenomenology when the system is subjected to two SBFS: a pre-existing pinning field and a stress applied as an experimental probe, and propose experimental signatures that may help distinguish a truly nematic superconducting order parameter from a single-component one.

In addition, in Section \ref{JosephsonSeciton} we consider the critical tunnel current in the Josephson junction between the proposed nematic superconducting state and another s-wave superconductor.  The idea of using the Josephson effect as a probe for unconventional pairing symmetry has long been considered\cite{Tanaka1997, Sumiyama1998, Tanaka1999, Stefanakis2002, Gouchi2012}.  The Ginzburg-Landau approach is also applicable to describe the Josephson tunneling\cite{Yip1990}.  If the nematic hypothesis does hold, the critical current shows a non-trivial anisotropic dependence on the SBF, as well as a temperature dependence that differs from the s-wave-to-s-wave scenario.  This provides another experimental test for the nematicity.

\section{Ginzburg-Landau Free Energy}
\label{GLsection}

%{\color{red}
We will formulate the GL theory by treating the system as ``almost $D_{3d}$-invariant''.  That is, we consider the nematic superconducting order parameter living in the $E_u$ representation of $D_{3d}$, and write down a $D_{3d}$-invariant free energy.  The pinning effect is introduced via coupling to a (possibly weak) explicit SBF.

This scenario is similar to the nematic pseudogap phase of YBCO\cite{Ando2002, Hinkov2008, Daou2010, Sato2017}, where the order parameter would have broken the four-fold rotation symmetry of the CuO$_2$ plane, except that this ``symmetry'' is already weakly broken by the orthorhombic lattice.  Nonetheless the nematic response is greatly enhanced by the nematic pseudogap order parameter.
%}

%\emph{(Is the next paragraph too combative?)}
%
%Hypothetically, if the explicit SBF is switched off, the electrons may exhibit a separate normal nematic phase just above the superconducting $T_c$\cite{Hecker2018}.  The full treatment of such a scenario requires a separate electronic nematic order parameter.  However, given the conflicting reports of normal state nematic response or the lack thereof\cite{Matano2016,Yonezawa2017,Willa2018,Sun2019,Smylie2018}, we do not feel that the inclusion of a separate nematic order parameter is warranted.  Furthermore, within the leading order GL expansion, this nematic order parameter is only to be treated as yet another background SBF near the superconducting $T_c$.  Our simpler theory should remain formally valid, and all qualitative predictions holds.
%

\subsection{Free Energy without SBF}

Let us first give a brief recap of the GL free energy for a nematic superconductor in the absence of any SBF.

We define the coordinate axes as follows: the $z$-axis is aligned with the principal $C_3$ axis of the lattice, the $x$-axis is along one of the $C_2$ axes in the basal plane, and the $y$-axis is chosen to form a right-handed set of coordinate system.

We assume that the superconducting order parameter is a complex two-component quantity $\vec{\eta}$.  It will be parameterized as:
\beq
\label{etaParam}
\vec{\eta} = 
	\bpm
		\eta_x \\
		\eta_y
	\epm
	= \sqrt{D}
	\bpm
		\cos \theta \\
		e^{i \phi} \sin \theta
	\epm.
\eeq
Here $\eta_x$ is kept real by a suitable global $U(1)$ phase rotation.    See Fig \ref{quintuple} for illustration.

The point group $D_{3d}$ can be built from three operations: $2\pi/3$-rotation about the $z$-axis (denoted $C_3$), $\pi$-rotation about the $x$-axis (denoted ${C_2}'$), and mirror reflection about the $yz$-plane (denoted $M$).  The odd-parity proposal\cite{Fu2014} would have $\vec{\eta}$ transforming in the $E_u$ representation, similar to $(x,y)$.

However, we may equally well consider the possibility where $\vec{\eta}$ transforms in the $E_g$ representation instead, similar to the pair $(yz, -xz)$.  Gauge invariance requires $\vec{\eta}$ and $\vec{\eta}^{*}$ to always come in pairs in the GL expansion, and the sign of parity is irrelevant.  The GL free energy and most of the subsequent analysis remain unchanged for the $E_g$ scenario, except that a minor modification is needed for the Josephson tunneling, which does not affect our qualitative conclusion.  For either case, one can consistently define $\eta_x$ to be the component invariant under ${C_2}'$.

The quantity $\vec{\eta}$ is to be considered as a headless vector, since  a sign in $\pm\vec{\eta}$ can be regarded as a $U(1)$ phase factor.  Therefore one may restrict $\theta$ to the interval $(-\pi/2, \pi/2]$, keeping in mind that $\theta$ and $\theta + \pi$ are always degenerate.

The GL expansion of free energy is obtained by writing down all the terms that are invariant under $D_{3d}$, global phase shift of $\vec{\eta}$, and time reversal.  Up to $O(\eta^6)$, the most generic form is:
\begin{widetext}
\beq
\label{freeEnergyNoStress}
\spl{
\fe_0 =& \; \alpha \vert \eta \vert^2
	+ \left [
		\beta_1' \vert \eta \vert^4
		+ \beta_2' \left (\vec{\eta} \cdot \vec{\eta} \right )
		\left ( \vec{\eta}^{\; *} \cdot \vec{\eta}^{\; *} \right) 
	\right ] \\
	&+ \left\lbrace
		\gamma_1' \vert \eta \vert^6
		+ \gamma_2' \left[
			(\eta_1 + i \eta_y)^3
			(\eta_x^{\; *} + i\eta_y^{\; *})^3
			+ (\eta_x - i \eta_y)^3
			(\eta_x^{\; *} - i\eta_y^{\; *})^3
		\right]
		+ \gamma_3' \vert\eta\vert^2
		\left( \vec{\eta}\cdot\vec{\eta} \right)
		\left ( \vec{\eta}^{\; *} \cdot \vec{\eta}^{\; *} \right)
	\right\rbrace.
}
\eeq
\end{widetext}
Here $\alpha$ is taken to be
\beq
\alpha = \kappa \left( \frac{T-T_0}{T_0}\right),
\eeq
where $T_0$ is the superconducting transition temperature without external field.  We ignore the temperature dependence of all other coefficients.

Let us first consider the free energy up to $O(\eta^4)$.  To this order $\fe_0$ is in fact invariant under arbitrary rotations in the $xy$-plane.  Two solutions for $\vec{\eta}$ are possible: the complex chiral state that is rotationally invariant but spontaneously breaks time-reversal symmetry, and the real nematic state which breaks the rotational symmetry but is time-reversal invariant.  The sign of $\beta_2'$ decides which phase is favored\cite{Volovik1988, Agterberg1995}.

Since the nematic state is by assumption the true solution, we take the appropriate sign $\beta_2' < 0$.  The minimum of $\fe$ now lies along $\phi = 0$, and in effect %
$\left (\vec{\eta} \cdot \vec{\eta} \right )\left ( \vec{\eta}^{\; *} \cdot \vec{\eta}^{\; *} \right) = \vert \eta \vert^4$.  Therefore we define $\beta \equiv \beta_1' + \beta_2'$, and require $\beta > 0$ for stability.

Now let us move on to include terms of $O(\eta^6)$.  The $\gamma_3'$ term essentially has the same $\theta$- and $\phi$- dependence as the $\beta_2'$ term.  Since a transition from nematic to chiral state at a lower temperature is never observed, we explicitly require $\gamma_3' < 0$.  On the other hand, it can be shown that all extrema of the $\gamma_2'$ term correspond to real values of $\vec{\eta}$.  Consequently, one may assume $\vec{\eta}$ is real and simply set $\phi = 0$.  In terms of $\theta$ and $D$, the free energy reads:
\beq
\fe_0 = \alpha D + \beta D^2
	+ \left[ \gamma_1 + \gamma_2 \cos \left( 6\theta \right) \right] D^3,
\label{freeEnergyNoStress2}
\eeq
where $\gamma_1 \equiv \gamma_1' + \gamma_3'$ and $\gamma_2 = 2\gamma_2'$.  For stability we require $\gamma_1 > 0$ and $\gamma_1 > \vert \gamma_2 \vert$.  

The $\gamma_2$ term breaks the full rotational symmetry down to a discrete six-fold symmetry.  This is the lowest order at which the crystal anisotropy enter the GL expansion\cite{Agterberg1995}.

\subsection{Stress as an SBF}

Stress serves as our prototypical SBF.  We will focus on stress in the $xy$-plane.  Generally, planar stress is a rank-two symmetric tensor $\hat{\ve}$ with three independent real parameters.  Under $D_{3d}$, this tensor decomposes into two parts: the scalar $\tr{\hat{\ve}}$, and the traceless part which is organized to form the two-component quantity
\beq
\label{stressParam}
\spl{
\vec{\ve} &=
	\bpm
		\ve_{xx} - \ve_{yy} \\
		- 2 \ve_{xy}
	\epm
\equiv
	\bpm
		\ve_1 \\
		\ve_2
	\epm\\
	&= \ve \bpm
		\cos \left[ 2 (-\Phi + \Phi_0) \right] \\
		\sin \left[ 2 (-\Phi + \Phi_0) \right]
	\epm,
}
\eeq
which transforms in the $E_g$ representation under $D_{3d}$.

%{\color{red}
Gauge invariance of the free energy requires that $\vec{\eta}$ and $\vec{\eta}^{\,*}$ to pair up, and all allowed combinations are parity-even.  Therefore to couple to the nematic superconductor at all, an SBF must transform under the even-parity $E_g$ representation.  It therefore shares the same phenomenology with stress \eqref{stressParam}.  One can simply replace the stress with any pre-existing SBF of an arbitrary origin, and the resulting GL free energy is formally identical.  In particular, we note that for externally applied \emph{strain}, the form of GL free energy \eqref{fullFreeEnergy} remains exactly identical.  The result in this paper equally applies to experiments which uses strain instead of stress.
%}

We choose to parameterize $\vec{\ve}$ as \eqref{stressParam} because $\Phi$ would correspond to the physical rotation angle about the $z$-axis.  When the stress is rotated by an arbitrary angle $\delta$, $\vec{\ve}$ changes via $\Phi \rightarrow \Phi + \delta$.  The angle $\Phi_0$ defines the direction relative to the $x$-axis that corresponds to $\Phi=0$, and we leave it unspecified for now, to be chosen for convenience later.  The angle $\Phi$ defines a special direction in the basal plane, and subsequently we may refer to it as the ``orientation'' of the stress.  This breaks the point group rotation symmetry.

For the sake of completeness, we state that one can form another $E_g$ pair using out-of-plane stress components: the pair $(\ve_{yz}, -\ve_{xz})$ can be used in place of \eqref{stressParam}.  Note that this is possible only for the trigonal $D_3d$ group.  If B and C sites in Fig \ref{quintuple} were to be identified, the resultant hexagonal $D_{6h}$ point group wouldn't have allowed the replacement.

For the remainder of this paper, we will always assume that $\tr{\hat{\ve}}$ is kept constant.  The dependence of GL coefficients on $\tr{\hat{\ve}}$ can therefore be entirely disregarded.

We will only consider coupling to $\vec{\ve}$ at linear order.  As discussed above, gauge invariance requires $\vec{\eta}$ and $\vec{\eta}*$ to comes in pairs that transform under the $E_g$ representation.  Up to order $O(\eta^4)$, the following is the exhaustive list of possible combinations:
\ali{
\label{S}
\vec{S} =&
	\; \bpm
		\vert \eta_x \vert^2 - \vert \eta_y \vert^2 \\
		-\eta_x \eta_y^{\, *} - \eta_y \eta_x^{\, *}
	\epm
	\equiv \bpm
		S_1\\
		S_2
	\epm
	\\
\vec{T} =& \; \vert \eta \vert^2 \, \vec{S}, \\
\vec{U} =&
	\; \bpm
		\left( \vert\eta_x\vert^2 - \vert\eta_y\vert^2 \right)^2 
			-\left(\eta_x \eta_y^{\,*} + \eta_y \eta_x^{\, *} \right)^2\\
		2 \left(\eta_x \eta_y^{\, *} + \eta_y \eta_x^{\, *} \right)
			\left( \vert \eta_x \vert^2 - \vert \eta_y \vert^2 \right)
	\epm.
}

The additional terms entering the free energy are
\beq
	\fe_{\ve} = - g_0' \, \vec{\ve} \cdot \vec{S}
			- g_1' \, \vec{\ve} \cdot \vec{T}
			+ g_2' \, \vec{\ve} \cdot \vec{U}.
\label{stressFreeEnergy1}
\eeq
The leading $\vec{\ve} \cdot \vec{S}$ term was considered in the literature\cite{Venderbos2016a, Willa2018}.  We go beyond quadratic order in $\vec{\eta}$ to study the interplay between the applied stress and crystal anisotropy.

All three terms in \eqref{stressFreeEnergy1} can be shown to be extremized by real values of $\vec{\eta}$.  Together with \eqref{freeEnergyNoStress}, every term in the free energy favors a real $\vec{\eta}$.  One is again allowed to set $\phi = 0$ from this point on.

Let us first look at $\vec{\ve} \cdot \vec{S}$.  Define $g_0 = \vert g_0' \vert > 0$.  By appropriately choosing $\Phi_0 = 0$ or $\pi/2$, this term can be written as
\beq
-g_0' \, \vec{\ve} \cdot \vec{S} = 
	- g_0 \, \ve D \cos \left( 2\theta - 2\Phi \right)
\label{g0Term}
\eeq

The free parameter $\Phi_0$ has been used up to fix the sign of $g_0$; now the $\vec{T}$ term must be taken as-is.  Define $g_1 = \pm g_1'$ with appropriate sign depending on the previous choice of $\Phi_0$, and one may write:
\beq
-g_1' \, \vec{\ve} \cdot \vec{T} = 
	-g_1 \, \ve D^2 \cos \left( 2\theta - 2\Phi \right).
\label{g1Term}
\eeq

Similarly, the $\vec{U}$ term becomes:
\beq
g_2' \, \vec{\ve} \cdot \vec{U} =
	- g_2 \, \ve D^2 \cos \left( 4\theta + 2\Phi \right),
\label{g2Term}
\eeq
with $g_2 = \pm g_2'$ appropriately defined to absorb the possible sign due to the choice of $\Phi_0$.

To sum up, the stress dependent part of GL free energy is
\beq
\spl{
\fe_{\ve} =& \, - g_0 \, \cos \left( 2\theta - 2\Phi \right) \ve D \\
	&\; -\left[ g_1 \cos \left( 2\theta - 2\Phi \right)
		+ g_2 \cos \left( 4\theta + 2\Phi \right)\right] \ve D^2.
}
\label{stressFreeEnergy2}
\eeq
The coefficient $g_0$ is made positive by appropriate choice of $\Phi_0$.  The other two coefficients $g_1$ and $g_2$ can be of either signs.

It can be seen that \eqref{freeEnergyNoStress2}, \eqref{g0Term}, \eqref{g1Term} and \eqref{g2Term} are all invariant under six-fold (improper) rotation about the $z$-axis.  Consequently, we are free to impose the restriction $\Phi \in (-\pi/6, \pi/6]$, which amounts to redefining the $x$-direction relative to the external stress.

The full expression of GL free energy $\fe$ employed in this paper is the sum of \eqref{freeEnergyNoStress2} and \eqref{stressFreeEnergy2}.  Let us also introduce another notation:
\beq
\fe = \fe_{0} + \fe_{\ve}
= a D + b D^2 + c D^3,
\label{fullFreeEnergy}
\eeq
where $a$, $b$ and $c$ are functions of $\alpha$, $\theta$, $\Phi$ and $\ve$.

\subsection{Limit on the Magnitude of Stress}

The explicit form of coefficient $b$ in $\fe$ is
\beq
b = \beta - g_1 \, \ve \cos(2\theta - 2\Phi)
	- g_2 \, \ve \cos(4\theta + 2\Phi).
\eeq
If we take this expression at its face value, for $\ve$ large enough, the minimum value of $b$ turns negative, and the transition from normal to superconducting state becomes first order.  

While not implausible, this stress-induced first order transition has not been observed in any material to our knowledge.  We therefore limit the range of $\ve$ in our theoretical investigation to avoid this regime.  The appropriate condition is:
\beq
\ve \ll \frac{\beta}{\vert g_1 \vert}\, , \,  \frac{\beta}{\vert g_2 \vert}.
\label{stressLimit}
\eeq

\subsection{Sign of $\gamma_2$}

In the absence of an SBF, up to order $D^2$ the free energy $\fe_0$ is symmetric under arbitrary rotation around the z-axis.  That is, $\eta_x$ and $\eta_y$ are completely degenerate.  The six-fold symmetric $\gamma_2$ term breaks the degeneracy between $\eta_x$ and $\eta_y$.  When $\gamma_2 < 0$, the ground state has non-zero $\eta_x$, while $\gamma_2 > 0$ results in non-zero $\eta_y$\cite{Agterberg1995}.

Using a two-band lattice model, Fu\cite{Fu2014} argued that $\eta_y$ is the correct superconducting order parameter for $\text{Cu}_x \text{Bi}_2 \text{Se}_3$, which corresponds to $\gamma_2 > 0$.  We will thus assume the positive sign for the remainder of the paper.  However, we point out that all the result obtained subsequently can be easily mapped to the case where $\gamma_2 < 0$, should it turn out that way in experiment.

Let $\Delta \theta = \theta - \Phi$, and eliminate all occurrences of $\theta$ in favor of $\Delta \theta$ in \eqref{freeEnergyNoStress2} and \eqref{stressFreeEnergy2}.  By shifting $\Phi \rightarrow \Phi + \pi/6$, one effectively reverses the signs of both $\gamma_2$ and $g_2$.

\section{Phase Diagrams}
\label{PhaseDiagramSection}

In this section we will describe the three possible phase diagrams, assuming the GL free energy \eqref{fullFreeEnergy}.  The reduced temperature $\alpha$, the magnitude of the stress $\ve$, and the orientation of the stress $\Phi$ will span the three axes of the phase diagrams.  The derivation and analysis of the features on these phase diagrams will be given in later sections.

We want to emphasize that any of the non-trivial structure of the phase diagram is a direct consequence of the nematic superconducting order parameter $\vec{\eta}$.  An s-wave superconductor coupled to an explicit SBF may show anisotropic response, but will only exhibit a single superconducting phase, with $T_c$ depending on $\vec{\ve}$ analytically.

First let us consider the stress-free case $\ve = 0$.  The superconducting transition takes place at $\alpha = 0$, and the order parameter is six-fold rotationally degenerate: $\theta = -\pi/6$, $\pi/6$, or $\pi/2$.  We will refers to these directions as the ``natural minima'' of the free energy.

Let's turn on the stress.  By assumption, the size of stress $\ve$ is such that $b$ remain positive for all values of $\theta$ and $\Phi$.  The normal-to-superconducting transition is then solely controlled by the coefficient $a$ in \eqref{fullFreeEnergy}:
\beq
a = \alpha - g_0 \, \ve \cos(2\theta-2\Phi).
\eeq
At finite stress, the transition occurs at
\beq
\alpha = \alpha_1(\ve) \equiv g_0\,  \ve,
\label{criticalAlpha1}
\eeq
and the order parameter is directed at $\theta = \Phi$.  We will refer to this as the \emph{upper transition}.

\subsection{$\Phi$ = 0}

We first examine the case when the direction of stress is fixed at $\Phi = 0$.  This is a plane in the full three-dimensional parameter space $(\Phi, \ve, \alpha)$.

\begin{figure}
	\begin{subfigure}{0.45\linewidth}
		\includegraphics[width=\textwidth]{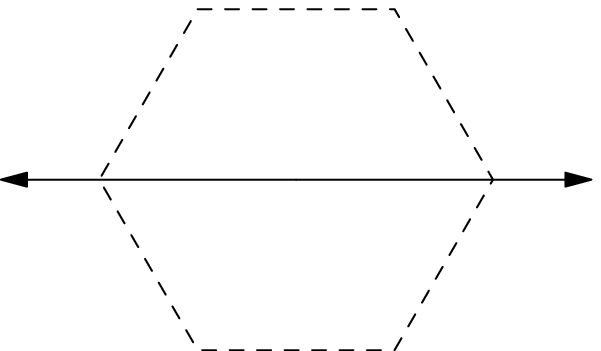}
		\caption{}
	\end{subfigure}
	\begin{subfigure}{0.45\linewidth}
		\includegraphics[width=\textwidth]{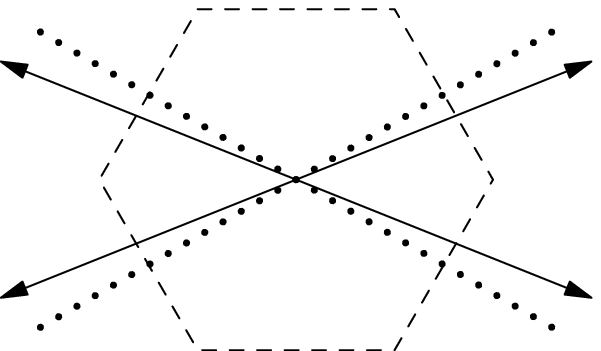}
		\caption{}
	\end{subfigure}
		\begin{subfigure}{0.45\linewidth}
		\includegraphics[width=\textwidth]{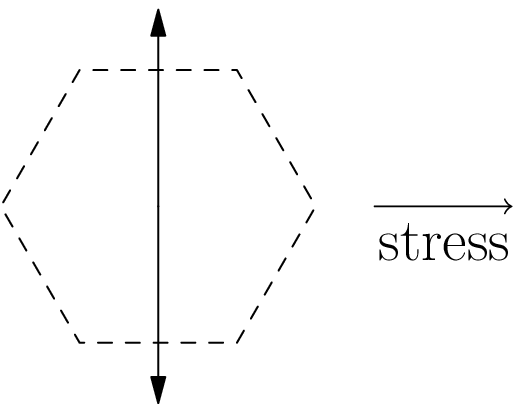}
		\caption{}
	\end{subfigure}
	\caption{\label{XPhases}The three allowed phases when the stress is oriented at $\Phi = 0$ as shown.  (a) Phase A: immediately below the upper transition.  (b) Phase B: between the middle and the lower transition if applicable. The two orientations depicted in phase B are degenerate and coexisting, and the dotted lines mark $\theta = \pm \pi/6$.  (c) Phase C below the possible lower transition.}
\end{figure}

We first orient the stress along $\Phi = 0$.  Below $T_c$, the order parameter is locked to $\theta=0$ by symmetry.  This is labeled as phase A.  As discussed earlier, the $\gamma_2$ term favors $\theta$ along one of the natural minima.  The competition between $g_0$ and $\gamma_2$ eventually leads to a second order phase transition from $\theta = 0$ to $\theta \neq 0$ at a lower temperature $\alpha_2(\ve)$.  This will be referred to as the \emph{middle transition}.  The lower temperature phase will be called phase B.  See Fig \ref{XPhases}.

%{\color{red}
This can be explained in more physical terms.  A hexagon has two types of high-symmetry directions.  When the stress favors the $x$-type direction, but the naturally preferred orientation is of the $y$-type (or vice versa), the middle transition exists as a result of the competition.
%}

This middle transition only exists for $\Phi = 0$.  Otherwise, $\theta$ will simply be pulled toward the nearer of $\pm \pi/6$ due to the reduced symmetry of the setup.  At $\ve = 0$, the upper and middle transition merge into a single superconducting transition.

Below $\alpha_2(\ve)$, the $\Phi = 0$ plane is a first order coexistent surface, separating the $\theta > 0$ and $\theta < 0$ phases on either side.  The middle transition is the critical end line of this surface.

Going still lower in temperature, another first order phase transition may occur at some $\alpha_3(\ve)$, and forms the lower bound of the coexistence surface.  It will be referred to as the \emph{lower transition}.  Below this lower transition, the orientation $\theta$ is fixed at $\pi/2$ (see FIG \ref{XPhases}).  The necessary and sufficient condition for the lower transition is
\beq
g_1 < g_2.
\label{lowerTransitionCriterion}
\eeq

The lower transition occurs at an temperature where $\vec{\eta}$ is no longer small.  The sixth order $\gamma_2$ term dominates the free energy, and by the same argument $g_0$ is overwhelmed by $g_1$ and can be ignored.  The $\gamma_2$ term has six degenerate minima (see equation \eqref{freeEnergyNoStress2},) and this degeneracy is lifted by $g_1$ and $g_2$.  It will be shown subsequently that the difference $(g_1 - g_2)$ determines which of the six minima is most favorable, leading to the  criterion \eqref{lowerTransitionCriterion}.

\begin{figure}
	\begin{subfigure}{0.48\linewidth}
		\includegraphics[width=\textwidth]{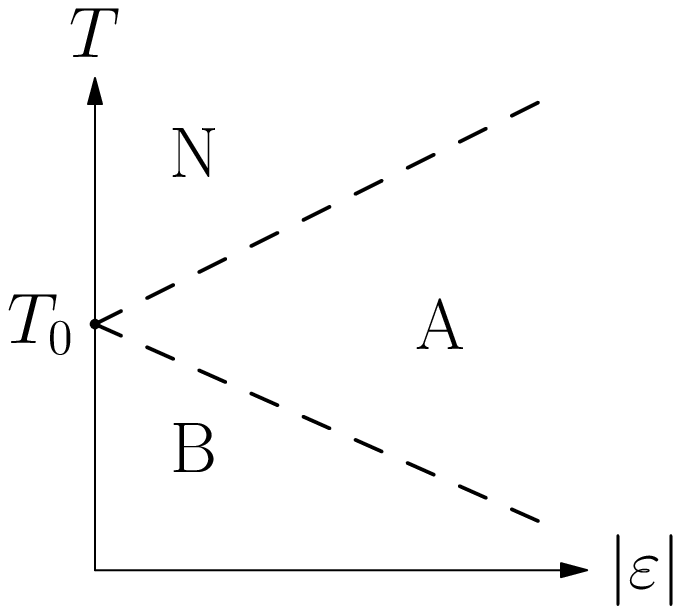}
		\subcaption{}
	\end{subfigure}
	\begin{subfigure}{0.48\linewidth}
		\includegraphics[width=\textwidth]{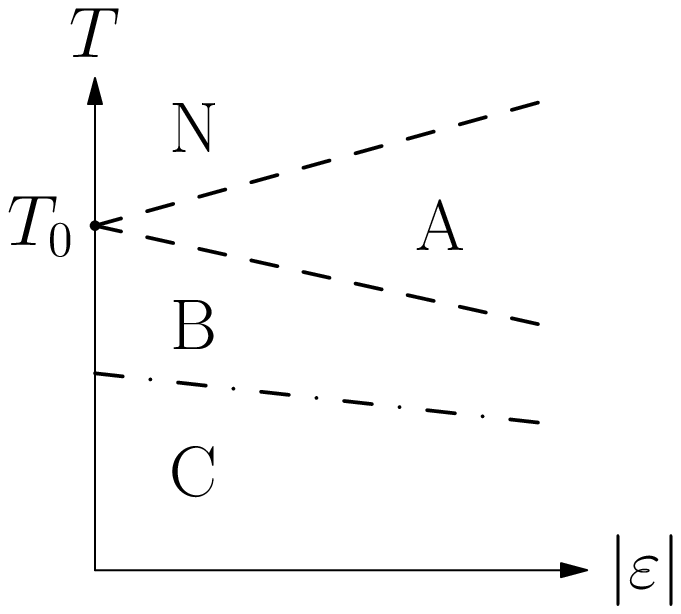}
		\subcaption{}
	\end{subfigure}
\begin{subfigure}{0.5\linewidth}
		\includegraphics[width=\textwidth]{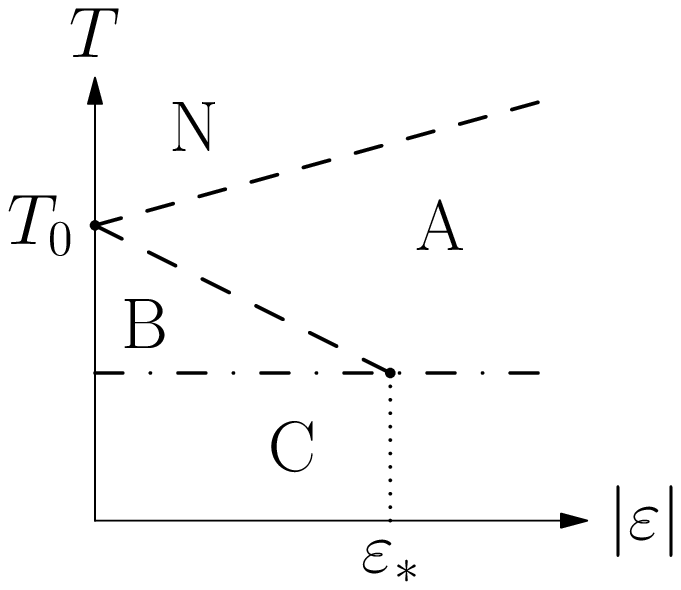}
		\subcaption{}
	\end{subfigure}
	\caption{\label{XPhaseDiagrams} The three possible phase diagrams when $\Phi = 0$.  The dashed lines denote second order transitions, and the dash-dotted lines denote first order transitions.  N denotes the normal phase.  (a) The lower transition is absent. (b) The lower transition is present.  (c) The lower transition is present, and the middle transition merges with it at $\ve = \ve_{*}$}
\end{figure}

All three lines of phase transition may continue indefinitely into higher stress, or alternatively the middle transition line may bend down and end when it merge with the lower transition at some value of stress $\ve_{*}$.  Overall, there are three different possible scenarios associated with this given $\fe$ at $\Phi = 0$, as shown in FIG \ref{XPhaseDiagrams}.

\subsection{$\Phi = \pi/6$}

Immediately below the upper transition the order parameter has $\theta = \pi/6$, which is again locked by the symmetry.  We shall refer to this as phase D.  This orientation also minimizes the $\gamma_2$ term, however, and there is no middle transition.

\begin{figure}
	\begin{subfigure}{0.48\linewidth}
		\includegraphics[width=\textwidth]{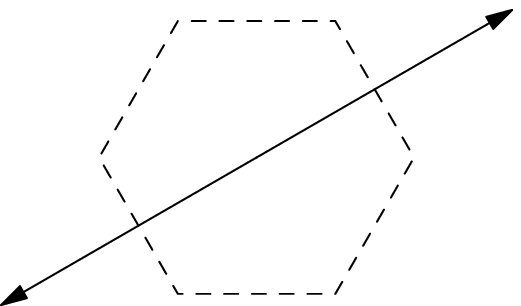}
		\caption{}
	\end{subfigure}
	\begin{subfigure}{0.48\linewidth}
		\includegraphics[width=\textwidth]{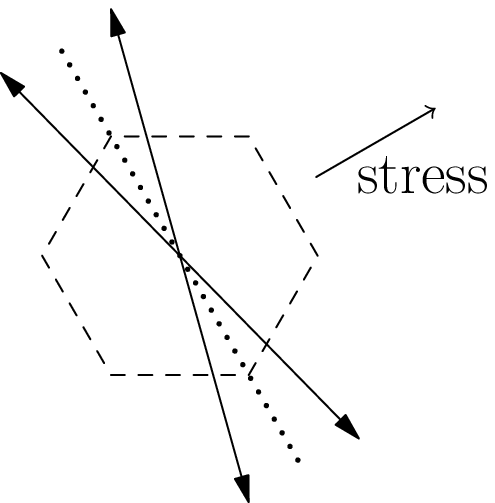}
		\caption{}
	\end{subfigure}
	\caption{\label{YPhases} The two allowed phases when the stress is oriented at $\Phi=\pi/6$. (a) Phase D: immediately below the upper transition.  (b) Phase E: the phase below the lower transition if it is present.  The dotted line marks the direction perpendicular to the stress.}
\end{figure}

On the other hand, the competition between $g_0$ and $g_2$ may be relevant at low temperature if \eqref{lowerTransitionCriterion} is satisfied, and then there is a lower transition.  Crossing this transition from high to low temperature, the equilibrium state goes from $\theta = \pi/6$ to two-fold degenerate $(\theta-\pi/6) > 0$ and $(\theta-\pi/6) < 0$.  Please see FIG \ref{YPhases} for illustration.

Purely from the symmetry standpoint, this transition can be of either first or second order.  It will be shown that, for stress $\ve$ smaller than some critical value $\ve_c$, the lower transition is of first order; beyond this point it becomes second order.  We will let $\alpha_4(\ve)$ denote the line of this lower transition.

Similar to the middle transition at $\Phi = 0$, lower transition here serves as the critical end line of a first order coexistent plane between the $\theta > \pi/6$ and $\theta < \pi/6$ phases on two sides of $\Phi = \pi/6$.

\begin{figure}
	\begin{subfigure}{0.45\linewidth}
		\includegraphics[width=\textwidth]{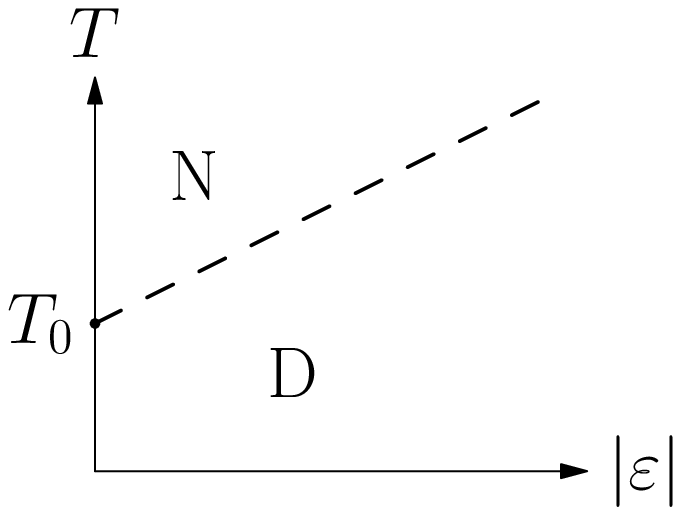}
		\caption{}
	\end{subfigure}
	\begin{subfigure}{0.45\linewidth}
		\includegraphics[width=\textwidth]{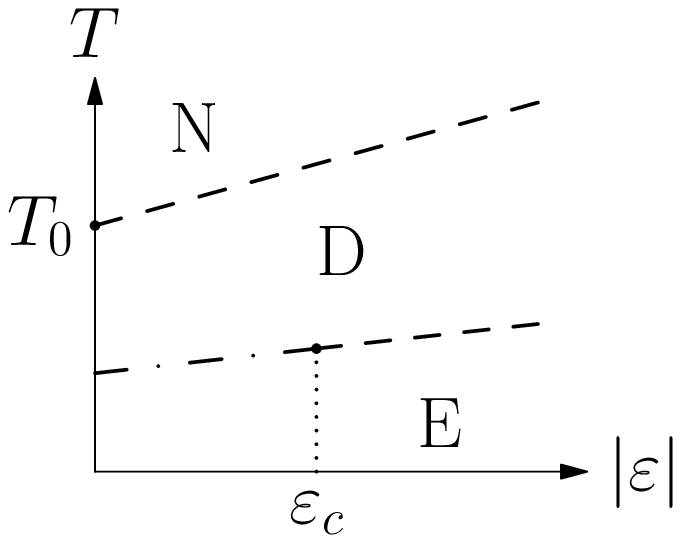}
		\caption{}
	\end{subfigure}
	\caption{\label{YPhaseDiagrams} The two possible phase diagrams when $\Phi = \pi/6$.  The dashed lines represent second order transitions, and the dash-dotted line represents first order transition.  N denotes the normal phase.  (a) the phase diagram in the absence of a lower transition; (b) the phase diagram in the presence of a lower transition.  The lower transition change from first to second order at $\ve = \ve_c$.}
\end{figure}

There are two possible scenarios at $\Phi = \pi/6$, one with and the other without a lower transition.  The phase diagrams are as shown in FIG \ref{YPhaseDiagrams}.

\subsection{The Full Phase Diagram}

\begin{figure}
	\begin{subfigure}{0.47\linewidth}
		\includegraphics[width=\textwidth]{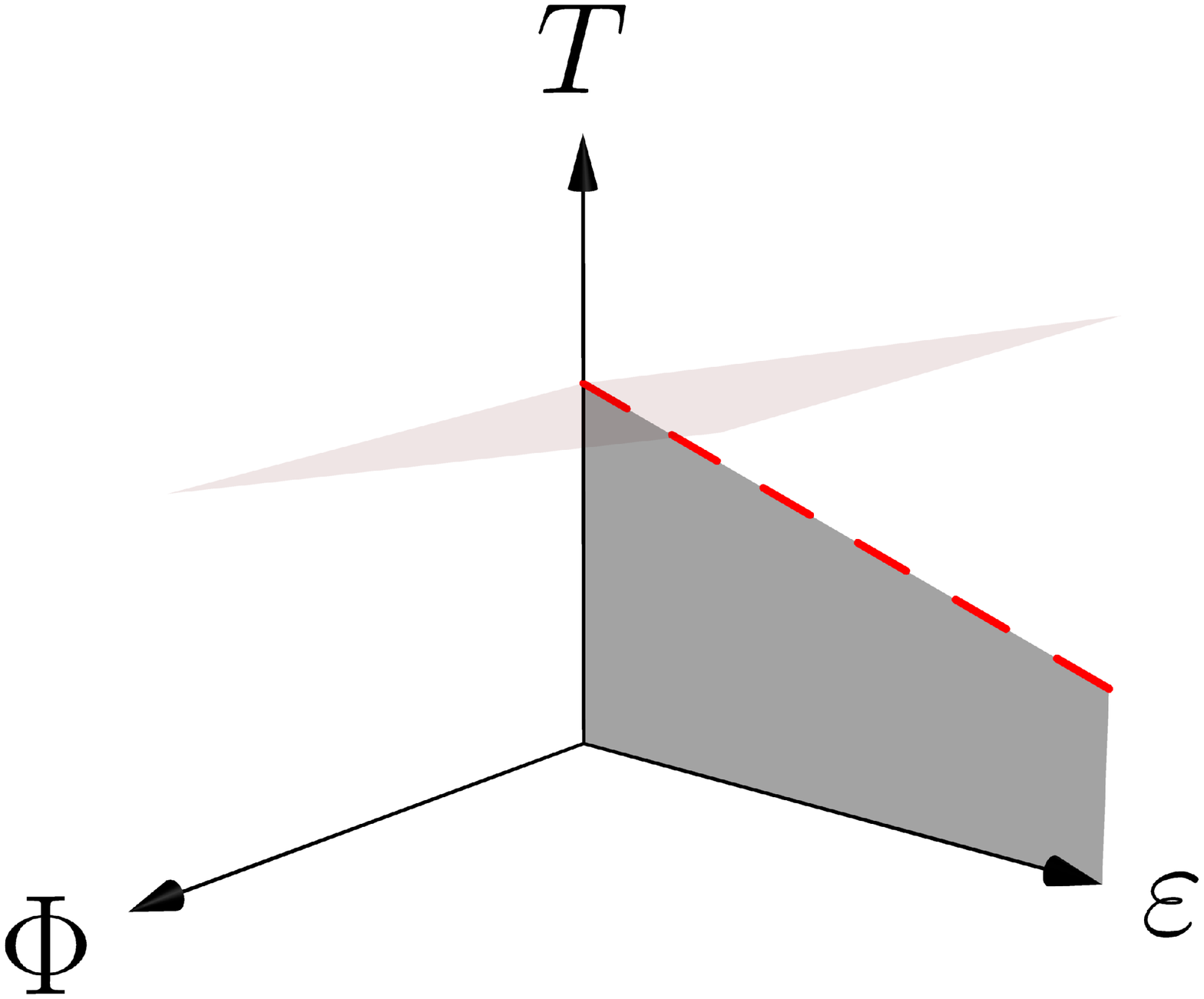}
		\caption{}
	\end{subfigure}
	\begin{subfigure}{0.47\linewidth}
		\includegraphics[width=\textwidth]{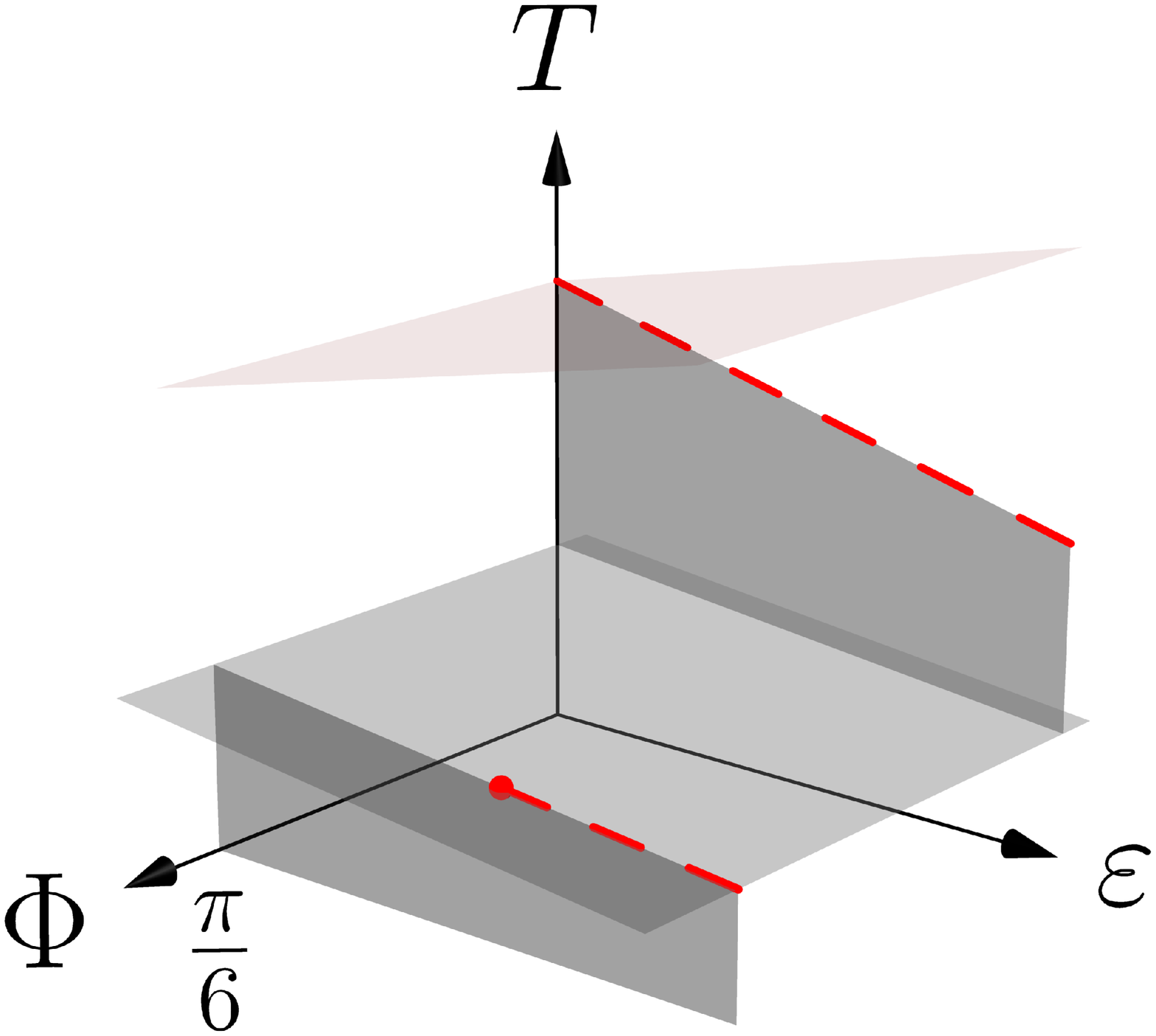}
		\caption{}
	\end{subfigure}
	\begin{subfigure}{0.47\linewidth}
		\includegraphics[width=\textwidth]{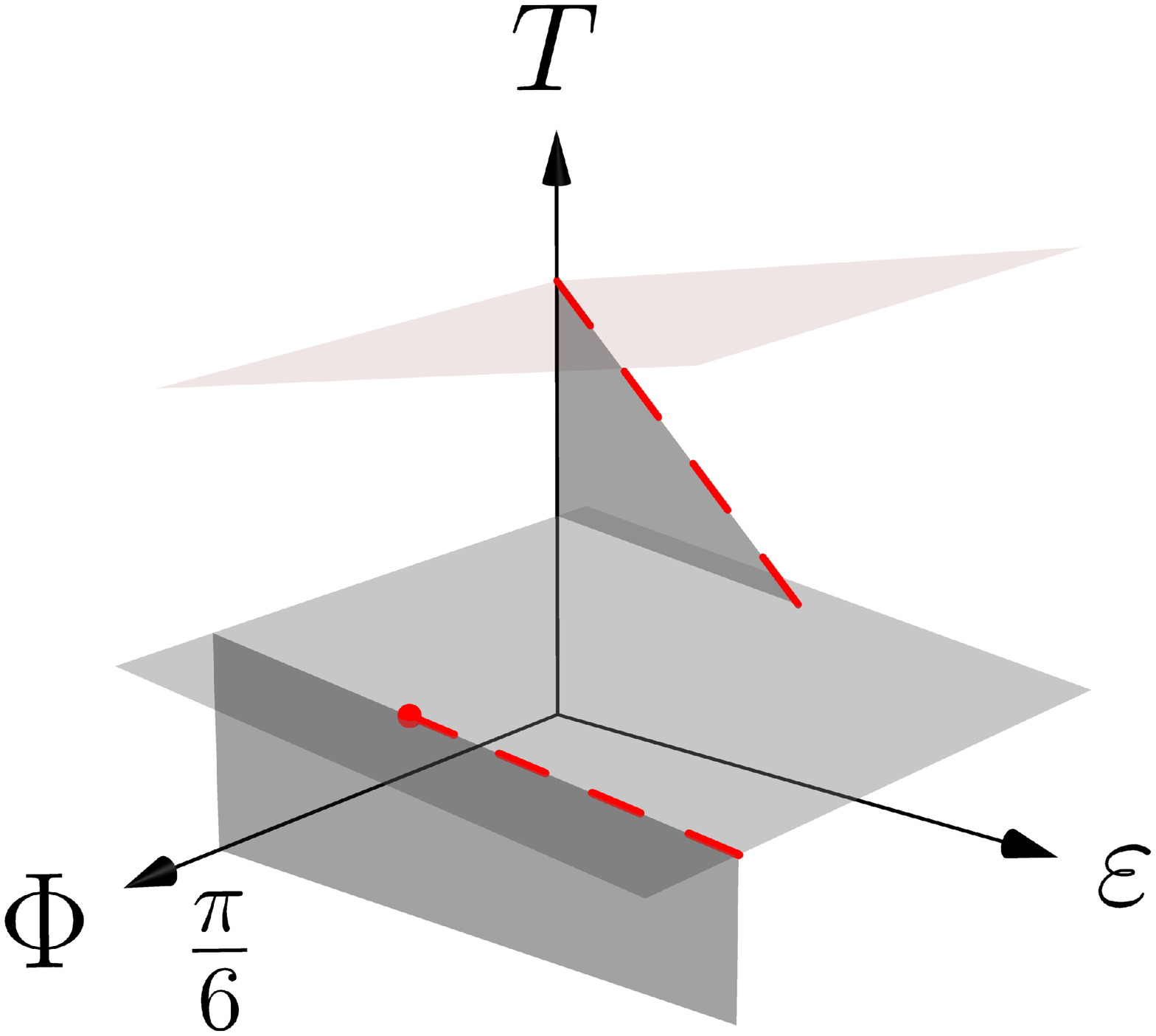}
		\caption{}
	\end{subfigure}
	\caption{\label{phaseDiagrams3d}  The possible phase diagrams, with the $\Phi$-axis added.  The upper shaded surface is the second order upper transition that separate normal and superconducting phases.  The other shaded surfaces are first order coexistence surfaces, while the thick dashed line represents second order critical end lines.  (a) Lower transition is absent.  (b) Lower transition is present.  (c) Lower transition is present, and the middle transition line ends on it.}
\end{figure}

A generic value of $\Phi$ breaks the six-fold rotational symmetry, and there cannot be a middle transition.  If the \eqref{lowerTransitionCriterion} is satisfied, the resulting lower transition indeed span a first order coexistent surface, interpolating between the lines of lower transition we have identified at $\Phi = 0$ and $\Phi = \pi/6$.  Overall there are three possible scenarios, as shown in FIG \ref{phaseDiagrams3d}.

As discussed earlier, the magnitude of stress in our discussion is limited by \eqref{stressLimit}.  Beyond the small-stress regime, the possible phase diagrams display very rich and complicated behaviors.  But as noted above, this regime is unlikely to be physically relevant.

The middle transition at $\Phi = 0$ marks the temperature below which $\gamma_2$ dominates over $g_0$.  The lower transition is due to the competition between $g_2$ and $\gamma_2$.  This already exhausts the list of possible competitions.  Consequently we do not anticipate any other transition on the phase diagrams.

We have numerically verified the assertions made in this section.  In the following sections, we will present analytical derivations the phase diagrams, and analyze their experimental implications.

\section{The Split Superconducting Transition}
\label{SplitTransitionSection}

\subsection{Upper Transition}

The normal-to-superconducting transition occurs at $\alpha_1(\ve)$, as given in \eqref{criticalAlpha1}.  Equivalently, one may revert to using the physical temperature:
\beq
T_1(\ve) = \kappa \, \alpha_1(\ve) + T_0
\label{criticalT1}
\eeq
is the corresponding critical temperature of the upper transition.  This critical temperature is isotropic and independent of $\Phi$.

However, the fact that $\alpha_1$ depends linearly in $\ve$ is itself an experimental signature for the nematic superconducting order.  A single-component order parameter cannot couple linearly to $\vec{\ve}$, as there is no way to form a combination that is invariant under three-fold rotation.

The specific heat jump will be anisotropic.  Following standard analysis, one recover
\beq
\Delta c_{v,1} = \frac{\kappa^2}{2 T_1(\ve)} \, 
	\frac{1}{\beta - \ve\,[g_1 + g_2 \cos(6\Phi)]}.
\label{cvJump1}
\eeq

This six-fold anisotropy is another unique signature of the nematic superconducting state.  It shares similar physical origin with the predicted six-fold anisotropy of $H_{c2}$ \cite{Agterberg1995a, Krotkov2002, Venderbos2016a}: the external field sees only the underlying lattice as the source of anisotropy.  By mapping out the $\ve$ and $\Phi$ dependence of $\Delta c_{v,1}$, one can in principle experimentally determine the Ginzberg-Landau coefficients involved.  In particular, the the relative size of $g1$ and $g2$ can be used to predict whether a lower transition may exist.

\subsection{Middle Transition}

The existence of a middle transition is another unique signature of the nematic superconducting state.  It separates the $\theta = 0$ phase from $\theta \neq 0$.  Therefore $\theta$ itself is an appropriate order parameter to describe this transition.  To this end, it is useful to first minizing $\fe$ with respect to variation in $D$.  Thus we define $\tilde{D}$ such that
\beq
\left( \frac{\partial \fe}{\partial D}\right)_{D = \tilde{D}} = 0.
\eeq
$\tilde{D}$ is an implicit function of $\theta$, $\alpha$, $\ve$, and $\Phi$.  We will assume that the stress orientation is tuned to $\Phi = 0$ for the remainder of this part.

One can now derive a Ginzberg-Landau expansion for $\theta$ by replacing $D$ with $\tilde{D}$:
\beq
\spl{
\tilde{\fe} &\equiv a \tilde{D} + b \tilde{D}^2 + c \tilde{D}^3 \\
&= k_0 + k_2 \, \theta^2 + k_4 \, \theta^4 + \dots
}
\eeq
Then the standard analysis of second order phase transition applies.

As noted in the previous section, the middle transition is due to the competition between $g_0$ and $\gamma_2$.  It can be shown that, for asymptotically small $\ve$, the leading expressions of coefficients $k_2$ and $k_4$ are independent of $g_1$ and $g_2$.  Here we present these asymptotic formulas, though the exact analytic result can be obtained by carefully retaining all terms.

The critical temperature $T_2$ is
\beq
T_2 = T_0 \left[
	1 - \frac{2}{3} \frac{\beta}{\kappa} \sqrt{\frac{g_0}{\gamma_2}}
		\sqrt{\ve}
	+ \frac{g_0}{\kappa}
	\left( \frac{2\gamma_2 - \gamma_1}{3\gamma_2}\right) \ve
\right],
\label{criticalT2}
\eeq
and the specific heat jump is
\begin{widetext}
\beq
\Delta c_{v,2} =
\frac{\kappa^2 T_2(\ve)}{T_0^2}
\frac{9 \,\gamma_2 \, g_0 \, \ve}
	{2 \left[
		g_0 \, \ve \, ( 4\gamma_1 - 5\gamma_2 + 4\beta \sqrt{\gamma_2
		\, g_0 \, \ve}) 
	\right]
	\left[
		\beta + (\gamma_1 + \gamma_2) \sqrt{g_0 \, \ve/\gamma_2} 
	\right]
	}.
\label{cvJump2}
\eeq
\end{widetext}

The formulas \eqref{criticalT2} and \eqref{cvJump2} are accurate in the limit where $(g_1 \, \ve /\beta)$, $(g_2 \, \ve/ \beta)$, $g_1 \sqrt{\ve / g_0 \, \gamma_2}$ and $g_2 \sqrt{\ve / g_0 \, \gamma_2}$ are small.

We also note that $\Delta c_{v,2}$ vanishes in the limit $\ve \rightarrow 0$.  This is consistent with the sum rule
\beq
\lim_{\ve \rightarrow 0} \left( \Delta c_{v,1} + \Delta c_{v,2} \right)
	= \Delta c_v,
\eeq
where $\Delta c_v$ is the specific heat jump of the superconducting transition at zero stress.

\section{Lower Transition}
\label{LowerTransitionSection}

The existence of a lower transition can be inferred by looking at the extremely low temperature $\alpha \rightarrow -\infty$ limit:
\beq
\spl{
\tilde{\fe} \approx& \;
\vert\alpha\vert^{3/2} \left[
	3\gamma_1 + 3\gamma_2 \cos(6\theta)
	\right] \\
& \; + \frac{
	\vert\alpha\vert
	\left\lbrace
		\beta - \ve\left[ g_1 \cos(2\theta - 2\Phi)
			+g_2 \cos(4\theta + 2\Phi)
		\right]
	\right\rbrace 	
	}
	{
	3\left[
	\gamma_1 + \gamma_2 \cos(6\theta)\right]
	}\\
& \; + O\left(\vert\alpha\vert^{1/2}\right).
}
\label{freeEnergyLowT}
\eeq
The leading term has the six-fold rotational symmetry, favoring all natural minima equally.  The sub-leading term lifts this degeneracy.

The sign of $(g_1-g_2)$ controls the qualitative behavior of \eqref{freeEnergyLowT}.  First assuming a generic value of $\Phi$ that breaks the six-fold rotational symmetry.  If $g_1 > g_2$, out of the three minima, the one \emph{closest} to $\Phi$ wins out.  Conversely if $g_1 < g_2$, the minimum \emph{furthest away} from $\Phi$ becomes the lowest.  On the other hand, the order parameter has $\theta = \Phi$ at the upper transition, and then initially drifts toward the nearest of the natural minima as the temperature is lowered.  For $g_1 < g_2$, there must be a first order transition separating the low-temperature asymptotic behavior from that just below the upper transition.  This justifies \eqref{lowerTransitionCriterion} as the criterion for a lower transition.

The cases of $\Phi = 0$ and $\pi/6$ require special attention.  For the sake of clarity, we introduce $\sigma = (\theta - \Phi)$, and consider the range $-\pi/2 < \sigma \leq \pi/2$.  It can be shown that $\tilde{\fe}$ exhibits the following properties at $\Phi = 0$ or $\pi/6$:
\begin{enumerate}
\item
$\tilde{\fe}$ is symmetric under $\sigma \rightarrow -\sigma$
\item
$\tilde{\fe}$ is always stationary at $\sigma = \pi/2$ and $0$.
\item
$\tilde{\fe}$ admits at most five stationary points within the range $-\pi/2 < \sigma < \pi/2$
\footnote{To see this, one notes that for a given $\sigma$, the relationship between $\tilde{D}$ and $\alpha$ is invertible.  Instead of finding an upper bound on number of stationary points along the path of constant $\alpha$ and $\ve$, one may solve for the much easier case of constant $\tilde{D}$ and $\ve$.
}
\item
Just below the upper transition, $\sigma = 0$ is the global minimum of $\tilde{\fe}$, and $\sigma =  \pi/2$ is the global maximum
\end{enumerate}

First let us look at the $\Phi = 0$ case.  Coming down from high temperature, $\sigma$ is initially fixed at $0$, but drifts away from this high-symmetry direction below the middle transition.  The previous argument for generic value of $\Phi$ therefore applies equally.  However, below the lower transition, the location of global minimum in this case is pinned \emph{exactly} at $\sigma = \pi/2$ by the enhanced symmetry.

Now we turn to the case of $\Phi = \pi/6$.  As noted in the previous section, the lower transition here may be either first or second order.

If the lower transition is first order, $\tilde{\fe}$ must develop a maximum-minimum pair on each side of $\sigma = 0$ as the temperature is lowered.  This already accounts for five stationary points within $-\pi/2 < \sigma < \pi/2$, and $\sigma = 0$ must always remain a local minimum.  On the other hand, a second order lower transition implies the stationary point at $\sigma = 0$ must revert its character as the temperature is lowered.  Therefore the criterion separating first and second order transition is whether $( \partial^2 \tilde{\fe}/\partial \sigma^2)_{\sigma = 0}$ changes sign when the temperature is lowered.  This gives the critical stress:
\beq
\ve_c = \frac{36 \, g_0 \gamma_3}{(g_1 - 4 g_2)^2}.
\label{criticalStressC}
\eeq
If $\ve > \ve_c$, the lower transition becomes second order.

Finally, we will address the question of if and when the middle transition at $\Phi = 0$ ends on the surface of lower transition.  While a closed-form analytic solution is possible, the full expression is extremely long and unwieldy.  We instead supply a recipe here.

The line of middle transition $\alpha_2(\ve)$ is implicitly defined by
\beq
\left( \frac{\partial^2}{\partial \sigma^2} \,
\tilde{\fe}\left(\alpha = \alpha_2, \ve, \sigma, \Phi = 0\right)
\right)_{\sigma = 0} = 0.
\eeq

Since the phase below the lower transition has exactly $\sigma = \pi/2$ at $\Phi = 0$, the equation
\beq
\spl{
\tilde{\fe}( \alpha_2(\ve_{*})&, \ve_*, \sigma = 0, \Phi=0 ) \\
&= \tilde{\fe}( \alpha_2(\ve_{*}), \ve_*, \sigma = \pi/2, \Phi=0 )
}
\eeq
determines the stress $\ve_{*}$ at which the middle and lower transitions meet if a solution exists.  If there is no solution, then the line of middle transition extends indefinitely into large $\ve$.  Our numerical results indicate that, depending on the actual values of GL coefficients, $\ve_{*}$ can be well within the limit \eqref{stressLimit}, and the ending of the middle transition can be physically relevant.

%{\color{red}
\section{Pre-existing Pinning Field and Signatures of Nematic Superconductivity}
\label{PinningFieldSection}

As noted in the introduction, there must be a pre-existing SBF in the $M_x$Bi$_2$Se$_3$ crystal to explain the pinning of the two-fold anisotropic direction.  Thus far, we have assumed that mechanical stress is the only SBF present in our GL analysis.  Should this be the case, one has direct control the SBF by application of another external stress, and the predictions in the previous sections can be directly tested.

However, the true nature of the pre-existing SBF is yet unknown, and it may not be mechanical stress or strain after all.  In this section, we consider the crystal subjected to both a \emph{pre-existing} pinning field $\vec{p}$ that is beyond our experimental control, and an \emph{artificially applied} stress $\vec{\ve}$ as an experimental probe.

The pinning field $\vec{p}$ must also transform under the $E_g$ representation.  In the same notation developed for stress, $\vec{p}$ can be expressed as
\beq
\vec{p} = p \bpm
			\cos (-2\Phi_p) \\
			\sin (-2\Phi_p)
			\epm
		\equiv \bpm
			p_1 \\
			p_2
			\epm
\label{pinning}
\eeq
where the magnitude $p$ is positive.  In the absence of stress or any other external SBFs, the orientation of $\vec{\eta}$ is pinned at $\theta = \Phi_p$ near the superconducting $T_c$.

As discussed in the introduction, the pinning effect is robust up to room temperature.  Therefore we find it reasonable to model the $\vec{p}$ as essentially a given background, independent of temperature, near the superconducting $T_c$ ($\approx 3$K).  While $\vec{p}$ may itself depends on the applied stress, one can formally expand
\beq
\vec{p}(\vec{\ve}) = \vec{p}(\vec{\ve} = 0) + \dots
\eeq
and the higher order terms merely renormailzes the GL coefficients.

To our knowledge, the reported two-fold anisotropy always aligns with a lattice direction in all known cases\cite{Pan2016,Yonezawa2017,Du2017, Smylie2018}.  We therefore conjecture that the pinning field, whatever it may be, breaks the lattice symmetry by favoring one of the three two-fold axes in the $\text{Bi}_2\text{Se}_3$ structure.

The two-fold axis of $H_{c2}$ in transverse field at zero applied stress can be identified experimentally.  We can redefine this two-fold direction as the $x$-axis.  Following the analysis of \cite{Venderbos2016a}, one deduces that the order parameter $\vec{\eta}$ is pinned at either the $(1,0)$ or $(0,1)$ direction.  The (unknown) sign of the GL coefficient to a certain gradient term determines the correct choice.  The two cases respectively yield:
\beq
\spl{
\vec{\eta} \propto  (1, 0), \quad
	\theta =&\; \Phi_p = 0, \quad
	\vec{p} \propto  (1,0); \quad \text{or} \\
\vec{\eta} \propto (0, 1), \quad
	\theta =&\; \Phi_p = \frac{\pi}{2}, \quad
	\vec{p} \propto (-1,0).
}
\label{conjectures}
\eeq

Now we consider how the system behaves when both the pre-existing SBF $\vec{p}$ and the applied stress $\vec{\ve}$ are present.  We will focus on the split superconducting transition here, i.e. the pair of upper and middle transitions.  To this end, one only needs to retain the coupling terms at order $O(\eta^2)$.  The modified GL coupling terms reads:
\beq
\fe_p = -\left( g_0 \vec{\ve} + g_p \vec{p}\right) \cdot \vec{S},
\eeq
where $S$ is given by \eqref{S}.

We will define the so-called ``total SBF'' in the bulk of the superconductor as
\beq
\vec{P}_b \equiv g_0 \vec{\ve} + g_p \vec{p},
\eeq
with the $\Phi_b$ the corresponding orientation angle.  The GL free energy under consideration is:
\beq
\fe = \left[\alpha - \vert P_b \vert \cos\left( 2\theta - 2\Phi_b \right) \right] D
		+ \beta D^2 + \dots
\eeq

The ratio $g_p/g_0$ is unknown.   However, one can always make $\vec{\ve} \propto (1,0)$ so that it is parallel to $\vec{p}$.  Physically, this amounts to keeping $\ve_2 = -2\ve_{xy} = 0$ while varying $\ve_1 = \ve_{xx}  - \ve_{yy}$.  The the change in magnitude $\Delta \vert P_b \vert$ is linearly dependent on the applied $\ve_1$.

\subsection{Upper transition}

Following \eqref{criticalAlpha1} and \eqref{criticalT1}, the critical temperature of the upper (normal-to-superconducting) transition will show a kink when one continuously varies the applied stress:
\beq
(T_1 - T_0) \propto \vert P_b \vert.
\label{TcPinningNematic}
\eeq
This behavior is a unique signature of a nematic superconducting state.  If the order parameter is single-component, it cannot couple linearly to either stress or the SBF.  One would expect instead a quadratic dependence:
\beq
(T_1 - T_0) \propto \vert P_b \vert^2.
\label{TcPinningSingleComponent}
\eeq
In either case, the minimum of $T_1$ also marks the point where the two-fold direction is tilted by $\pi/2$.

If the kink in \eqref{TcPinningNematic} can be identified, the predicted six-fold anisotropy purely due to the lattice should be restored at that point.  This effect can be seen from the specific heat jump, following \eqref{cvJump1}.  Following \cite{Venderbos2016a}, the upper critical field $H_{c2}$ should also exhibit the same six-fold anisotropy.  This provides further verification for the nematic state.

\subsection{Middle Transition}

The existence of the middle transition is another unique signature of the nematic superconducting state.  Recall that it is due to the competition between the naturally preferred alignment of $\vec{\eta}$ and the explicit SBF.

We have been assuming that $\eta_y$ is the naturally preferred order parameter, or that the GL coefficient $\gamma_2$ is positive.  The middle transition exists when the orientation of the total SBF obeys $\Phi_b = 0$, or $\vec{P}_b \propto (1, 0)$, which is satisfied on only one side of the $\vert P_b \vert = 0$ kink.  The second order phase transition may in principle be identified by the jump in specific heat \eqref{cvJump2}.

We have made two assumptions here: the natural preferred direction, and the orientation of the pre-existing pinning field \eqref{conjectures}.  First, it may well be that $\eta_x$ is the natural preference instead, but the conclusion that the middle transition exists on only one side of the $\vert P_b \vert = 0$ kink stands unaltered.

It is trickier if the pre-existing $\vec{p}$ does not in fact align with a lattice direction.  Should this turns out to be the case, one can instead apply both non-zero $\ve_1$ and $\ve_2$, keeping the ratio $\ve_1/\ve_2$ constant.  The $\Phi_b = 0$ condition is satisfied by exactly one value of the magnitude $\ve$, and the middle transition exists only at that point.  Nevertheless, just off that exact value, the sharp middle transition is smeared into a crossover, and one can still register a steep increase in specific heat as an experimental signature.

%}

\section{Josephson Junction with an S-wave superconductor}
\label{JosephsonSeciton}

In this section, we change gear slightly and discuss the phenomenology of tunneling current between a nematic superconductor and an s-wave superconductor.  As pointed out by Yip \emph{et. al.}\cite{Yip1990}, one may write down the effective Hamiltonian of a junction purely based on symmetry considerations, in the same spirit of the GL theory.  The behavior of the critical current then reflects the symmetry of the nematic order parameter.  Here the implicit assumption is that the interface itself has spin-orbit coupling\cite{Millis1988} so as to allow a tunneling supercurrent.

The theory for Josephson tunneling between two s-wave superconductors was worked out in the classic paper by Ambegaoker and Baratoff\cite{Ambegaokar1963} (AB).  For a junction between two different s-wave superconductors, right below the lower of the two critical temperatures, the critical current is proportional to $\sqrt{T_c-T}$, and shows no strong stress dependence.  In this section we explore how the proposed nematic superconducting state gives rise to qualitative differences.

We consider the scenario where an s-wave superconductor (with a much higher $T_c$) is attached to the $M_x\text{Bi}_2\text{Se}_3$ sample being tested.  The surface contact of the sample is cut perpendicular to the $z$-axis.  We assume that the junction is in the tunneling limit.

%{\color{red}
The two-fold rotation ${C_2}'$ (and space inversion) in the original $D_{3d}$ group is no longer a valid symmetry on the surface of contact.  We will denote the reduced symmetry group as $G$.  The junction coupling term in the effective Hamiltonian respects $G$, and of course the $U(1)$ gauge symmetry.
%}

The s-wave superconductor is described by a complex scalar order parameter $\Psi = e^{i\chi} \vert \Psi \vert$.  We make the gauge choice so that the order parameter $\vec{\eta}$ is real, and $\chi$ represents the phase difference between the two superconductors.

%{\color{red}
\subsection{The Intrinsic Contribution}

Let us first discuss the leading order junction term that is independent of any SBF.  One can form $G$-invariant combinations at order $O(\eta^3)$, but these are not gauge-invariant in their own right, and must be coupled to $\Psi$.  And then the time-reversal symmetry demands the coupling terms to be overall Hermitian.  The leading coupling term, at $O(\Psi)$ and $O(\eta^3)$, has the form:
\beq
\spl{
\fe_{j0} =&\; m_0 \Psi^{*} \left( \eta_x \, S_2 - \eta_y \, S_1 \right) + \text{h.c.} \\
=& \; m_0 \Psi^{*} (\vec{\eta} \wedge \vec{S}) + \text{h.c.},
}
\label{junction0}
\eeq
where $S$ is defined in \eqref{S}.  The coefficient $m_0$ must be real to satisfy time reversal symmetry.

Here, we assume that the junction is but a small perturbation to the bulk, and that the system is sufficiently close to the upper transition.  One may therefore use the unperturbed solution for the nematic order parameter $\vec{\eta}$.  Then the coupling \eqref{junction0} reduces to
\beq
\fe_{j0} = - m_0 \vert\Psi\vert D^{3/2} \, \sin \left( 3\theta \right) \cos(\chi),
\eeq
where $\theta$ is the orientation of the order parameter $\vec{\eta}$ in the bulk.

The supercurrent across the junction can be identified\cite{Yip1990} as $2 \, (\partial \! \fe_{j0}/\partial \chi)$.  The factor of two is due to that the order parameters in the GL theory describe Cooper pairs.  Use also the fact that $D \propto \vert\alpha\vert$ just below the upper transition, one identify the critical current across the junction:
\beq
I_{c0} \propto \left(T_1 - T \right)^{3/2} \vert\sin 3\theta \vert.
\label{Ic0}
\eeq

In the hypothetical case where no SBF is present, \eqref{Ic0} is the only contribution to the critical current.  Our default scenario is that $\eta_y$ is naturally preferred, or that $\theta = \pi/2$ here.  Then \eqref{Ic0} shows a $3/2$ power-law temperature dependence that is markedly different from the AB theory.  The other scenario is that $\theta = 0$, and tunneling is completely disallowed, again a drastic departure from the AB theory.  When the total SBF in the bulk $\vec{P}_b$ is non-vanishing, $\theta = \Phi_b$ in the immediate vicinity of the upper transition.

\subsection{The SBF-assisted Contribution}

The quantity $S$ can be replaced by any SBF to yield an invariant coupling term.  Here we will impose both a pre-existing $\vec{p}$ and an applied stress $\vec{\ve}$.  The coupling term is thus
\beq
\fe_{j1} = \Psi^{*} \, \vec{\eta} \wedge
	\left( m_{p}\vec{p} + m_{\ve} \vec{\ve}\right) + \text{h.c.}
\label{junction1}
\eeq
Again, $m_p$ and $m_\ve$ are real by time reversal invariance.

Unfortunately, the ratio $m_{p}/m_{\ve}$ is not necessarily the same as $g_0/g_{\ve}$.  One defines another ``total'' SBF for the junction:
\beq
\vec{P}_{j} \equiv \; m_p \, \vec{p} + m_{\ve} \, \vec{\ve},
\eeq
which is in general not aligned with $\vec{P}_b$ in the bulk.

Let $\Phi_j$ be the orientation angle corresponding to $\vec{P}_j$.  Again using the bulk solution for $\vec{\eta}$, \eqref{junction1} becomes
\beq
\fe_{j1} = - \Psi^{*} \vert P_j \vert \sqrt{D} \, \sin ( \theta + 2\Phi_j ) \cos \chi.
\label{junction1-2}
\eeq

For a non-zero $\vert P_j \vert$, \eqref{junction1-2} is dominant in the immediate vicinity of the upper transition, where $D$ is still small.  Taking this limit, the critical current is
\beq
I_{c1} \propto \vert P_j \vert \left( T_1 - T \right)^{1/2} \vert \sin (\theta + 2\Phi_j) \vert.
\label{Ic1}
\eeq

If $\vec{\ve}$ and $\vec{p}$ aren't aligned, the angles $\Phi_b$ and $\Phi_j$ are different in general.  This critical current therefore shows a two-fold anisotropy as one rotates the stress, and has the conventional square-root dependence on temperature.  At first sight, this tunneling current may seem indistinguishable from that of the anisotropic s-wave scenario, but there are still some dramatic signatures that are direct consequences of the nematic superconductivity.

\subsection{Experimental Signatures}

At zero applied stress and non-vanishing pre-existing pinning field, one has $\theta = \Phi_b = \Phi_j = \Phi_p$.  We proceed to discuss the two cases in our conjecture \eqref{conjectures} separately.  

If $\Phi_p = 0$, then at zero applied stress, both $I_{c0}$ and $I_{c1}$ vanishes, and there is no tunneling current allowed.  An applied stress at a generic direction tilts both $\theta = \Phi_b$ and $\Phi_j$, and turns on the tunneling current.

Alternatively, one keeps $\ve_2 = 0$ and applies a non-zero $\ve_1$ as proposed in the previous section.  $\Phi_b$ and $\Phi_j$ go from $0$ to $\pi/2$ as $\vec{P}_b$ and $\vec{P}_j$ crosses zero, respectively.  While this change in $\Phi_j$ does not affect the critical currents, the change in $\Phi_p$ switches on the tunneling current from zero.

On the other hand, if $\Phi_p = \pi/2$, tunneling current without applied stress is initially non-zero.  One again applies $\ve_1$ while keeping $\ve_2 = 0$.  When $\vec{P}_b$ crosses zero, the tunneling current is switched off.

Additionally, if one finds a region where $\vec{P}_j \approx 0$ while $\Phi_b = \pi/2$, then $\fe_{j1}$ is anomalously suppressed by the fine-tuning of the applied stress, and the critical current will instead show the $3/2$ power-law temperature dependence of $I_{c0}$.

The above treatment amounts to the direct coupling of the two superconducting bulks.  This is admittedly an over-simplification: for unconventional superconductors, surface depairing may occur depending on the exact detail of the gap function\cite{Yip1996}.  The extent of this surface effect is dictated by the coherence length.  Therefore, near $T_c$ when the coherence length is large, surface depairing may substantially suppress the tunneling current; the overall temperature dependence will then have a higher power\cite{Tanaka1997,Yip1997} than our simple prediction of $3/2$.  Nonetheless, the temperature dependence is clearly distinct from the s-wave-to-s-wave case.

\subsection{$\vec{\eta}$ in $E_g$ Representation}

As advertised in the introduction, if $\vec{\eta}$ is instead even under space inversion, i.e. in the $E_g$ representation, the results in this section need minor modifications.  This stems from the fact that $M$, the mirror reflection about the $yz$-plane, acts differently on $E_u$ and $E_g$.  The coupling terms $\fe_{j0}$ and $\fe_{j1}$ becomes:
\beq
\spl{
{\fe_{j0}}' =& \; m_0 \Psi^{*} (\vec{\eta} \cdot \vec{S}) + \text{h.c.} \\
{\fe_{j1}}' =& \; \Psi^{*} \, \vec{\eta} \cdot
	\left( m_{p}\vec{p} + m_{\ve} \vec{\ve}\right) + \text{h.c.}.
}
\eeq
The upshot is that sines are to be replaced by cosines in both $I_{c0}$ and $I_{c1}$:
\beq
\spl{
{I_{c0}}' \propto & \; \left(T_1 - T \right)^{3/2} \vert\cos 3\theta \vert \\
{I_{c1}}' \propto & \; \vert P_j \vert \left( T_1 - T \right)^{1/2} \vert \cos (\theta + 2\Phi_j) \vert.
}
\eeq

The experimental signatures proposed above are still available, but all angles involved are shifted by $\pi/2$.  Without an applied stress $\vec{\ve}$, now tunneling current is forbidden at $\Phi_p = \pi/2$, and a non-zero $\vec{\ve}$ that tilts $\vec{P}_j$ away switch on the current.  On the other hand, $\Phi_p = 0$ initially allows a tunneling current that can be switched off by reversing $\vec{P}_j$.  The temperature dependence of $I_{c0}$ can be observed by having $\Phi_b = 0$ and $\vert P_j \vert \approx 0$.

%}

\section{Conclusion}

In this paper, we explore the coupling between the two-component superconducting order parameter $\vec{\eta}$ and the (traceless part of) stress $\vec{\ve}$ in the basal plane, and map out possible phase diagrams allowed by symmetry constraints.  Indeed the analysis is not restricted to mechanical stress: any SBF must couple in a similar manner.  In later sections we consider the the case where there are a pre-exisiting pinning SBF in the sample, and an applied stress as an experimental probe.

In the presence of a SBF, we found that the superconducting transition splits into two (\emph{upper} and \emph{middle} transitions), and there may be another phase transition at an even lower temperature (\emph{lower} transition), depending on the values of GL coefficients.

The critical temperature \eqref{criticalT1} of the upper transition does not show any anisotropy, but a \emph{six-fold} anisotropy for the specific heat jump \eqref{cvJump1} is predicted as the SBF is rotated.  The physics behind is similar to the theoretical angular dependence of upper critical field discussed in the literature\cite{Agterberg1995a, Krotkov2002, Venderbos2016a}.

The middle transition only exist when the SBF is aligned along $\Phi = 0$; the sharp transition becomes a crossover if the alignment is not exact.  We give formulas for the transition temperature \eqref{criticalT2} and specific heat jumps \eqref{cvJump2}.  As the middle transition is connected to the superconducting transition without a symmetry breaking field, we expect that these results from our GL theory are still quantitatively accurate.

The same cannot be said for the lower transition, which is not connected to the superconducting transition without a SBF.  The assumption that GL coefficients are temperature-independent may no longer be a quantitatively accurate approximation, and we focus our effort on the qualitative results.  The existence of a lower transition hinges on the criterion \eqref{lowerTransitionCriterion}.  It is first order almost everywhere, except when $\Phi = \pi/6$ and $\ve > \ve_c$ it becomes second order.  The critical value $\ve_c$ is given in \eqref{criticalStressC}.

%{\color{red}
The observed two-fold anisotropy of $H_{c2}$, and the robust pinning of the two-fold direction over many cycles, suggest that a pre-existing pinning field explicitly breaks the rotational symmetry in the sample.  We discuss how, at leading order in GL theory, the pinning field and an externally applied stress combine to form a ``total'' SBF felt by the sample.  By varying the applied stress, one gains access to different regions of the three-dimensional phase diagram worked out above.
%}

We point out two thermodynamics experimental signatures unique to the nematic state.  First, the superconducting critical temperature is \emph{linearly} proportional to the strength of the total SBF, as given in \eqref{TcPinningNematic}, as opposed to the quadratic relation if the order parameter is single-component.  Second, the existence of the middle transition, and the associated finite crossover even if the total SBF isn't exactly tuned to the right orientation, can be observed through calorimetry experiments.

%{\color{red}
When linked to another s-wave superconductor, the Josephson tunneling current also offers hints to the superconducting pairing symmetry.  We discuss how an applied stress can switch the tunneling current on and off if the superconducting order parameter is indeed nematic.  The unusual $(T_1 - T)^{3/2}$ temperature dependence of critical current $I_{c0}$ \eqref{Ic0} may also be seen in experiment if the required conditions are met.  None of these peculiar behaviors can be seen if the superconductivity of $M_x$Bi$_2$Se$_3$ turns out to be s-wave.
%}

We hope our findings here will guide future experimental effort in discerning the pairing symmetry of the $\text{Bi}_2\text{Se}_3$ family of superconductor, thereby helping to settle the debate on the topological nature of the superconductivity.

\begin{acknowledgments}
The authors would like to thank Anne de Visser for sharing the details of experimental procedures carried out in Pan \emph{et. al}\cite{Pan2016}, and Matthew Smylie for discussing the robustness of pinning effect seen in experiments.  This work is supported by Ministry of Science and Technology, Taiwan under grant number MOST 107-2112-M-001-035-MY3, and PTH is supported under grant number MOST 107-2811-M-001-045.
\end{acknowledgments}

\bibliography{nematicSC}

\end{document}